\documentclass[aps,prl,twocolumn,groupedaddress]{revtex4-2}

\usepackage{color}
\usepackage{xcolor}

\usepackage[utf8]{inputenc}
\usepackage[english]{babel}
\usepackage{amsmath,graphicx,enumerate}

\begin{document}


\title{Bubble phase induced by odd interactions} 

\author{L. Caprini$^{1}$}
\author{U. Marini Bettolo Marconi$^{2}$}
\affiliation{$^1$ Physics Department, Sapienza University of Rome, Piazzale Aldo Moro 5, Rome, Italy }
\affiliation{$^2$ Physics Department, University of Camerino, Via Madonna delle Carceri, Camerino, Italy }

\newcommand{\beq}{\begin{equation}}
\newcommand{\eeq}{\end{equation}}
\newcommand{\bea}{\begin{eqnarray}}   
\newcommand{\eea}{\end{eqnarray}}

\newcommand{\br}{{\bf r}}
\newcommand{\bu}{{\bf u}}
\newcommand{\bv}{{\bf v}}
\newcommand{\bR}{{\bf R}}
\newcommand{\bRz}{{\bf R}^0}
\newcommand{\bk}{{ \bf k}}
\newcommand{\bx}{{ \bf x}}
\newcommand{\vv}{{\bf v}}
\newcommand{\bn}{{\bf n}}
\newcommand{\mb}{{\bf m}}
\newcommand{\bq}{{\bf q}}
\newcommand{\bK}{{\bf K}}
\newcommand{\rb}{{\bar r}}
\newcommand{\rr}{{\bf r}}
\newcommand{\eb}{{\bf e}}

\newcommand{\kk}{\boldsymbol{\kappa}}
\newcommand{\greeketabold}{\boldsymbol{\eta}}
\newcommand{\xxi}{\boldsymbol{\xi}}
\newcommand{\cchi}{\boldsymbol{\chi}}
\newcommand{\bomega}{\boldsymbol{\Omega}}

\date{\today}


\begin{abstract}

We study a repulsive thermal system governed by odd interactions. The interplay between oddness and inertia induces a non-equilibrium phase transition from a homogeneous to a non-homogeneous phase, characterized by bubbles induced by odd interactions. This phenomenon occurs in the absence of attractions and is generated by the competition between the standard pressure contribution due to particle repulsion and an effective surface tension generated by odd-induced centrifugal forces. 
As a signature of the phase transition, the system exhibits vortex structures and oscillating spatial velocity correlations, which emerge close to the analytically predicted transition point. Our findings can be verified in granular experiments governed by odd interactions, such as spinners and colloidal magnets, and could be key to characterizing the emerging properties of metamaterials.
\end{abstract}

\maketitle

 Over the past decade, there has been a remarkable surge of interest in the properties of fluids with odd viscosity, solids with odd elasticity~\cite{fruchart2023odd,avron1998odd, banerjee2017odd, markovich2021odd, lou2022odd, reichhardt2022active, hosaka2023lorentz, hosaka2023hydrodynamics, everts2024dissipative, markovich2024nonreciprocity,scheibner2020odd,fruchart2020symmetries, alexander2021layered, ishimoto2022self, shankar2022topological,surowka2023odd, kobayashi2023odd}, and other systems with general odd transport coefficients~\cite{hargus2021odd, kalz2022collisions, kalz2024oscillatory, vega2022diffusive, poggioli2023odd}.
 Unlike standard materials, odd systems are characterized by elastic or viscosity tensors that include an antisymmetric contribution in the Navier or Navier-Stokes equations. Consequently, these systems do not conserve energy, violate time-reversal symmetry, and remain far from equilibrium~\cite{fruchart2023odd, yasuda2022time, epstein2020time}.
 To produce such unusual macroscopic properties, odd materials must consist of out-of-equilibrium microscopic units whose  interactions break parity symmetry. Odd systems are often called chiral~\cite{liebchen2022chiral} and include spinning particles, also known as spinners~\cite{scholz2018rotating, workamp2018symmetry, yang2020robust, lopez2022chirality}, which self-rotate and transversally transfer energy. Examples include granular systems where transverse pairwise interactions arise from friction~\cite{tsai2005chiral, han2021fluctuating}, spinning colloids where these forces are induced by the flow field advection generated by particle rotations~\cite{massana2021arrested, lenz2003membranes}, and colloidal magnets spun by a magnetic field~\cite{soni2019odd, mecke2023simultaneous}. Odd properties have also been observed in biological organisms, such as large assemblies of starfish embryos~\cite{tan2022odd} and the flagella of Chlamydomonas and sperm cells~\cite{ishimoto2023odd}.

\begin{figure}[!t]
\centering
\includegraphics[width=1\linewidth,keepaspectratio]{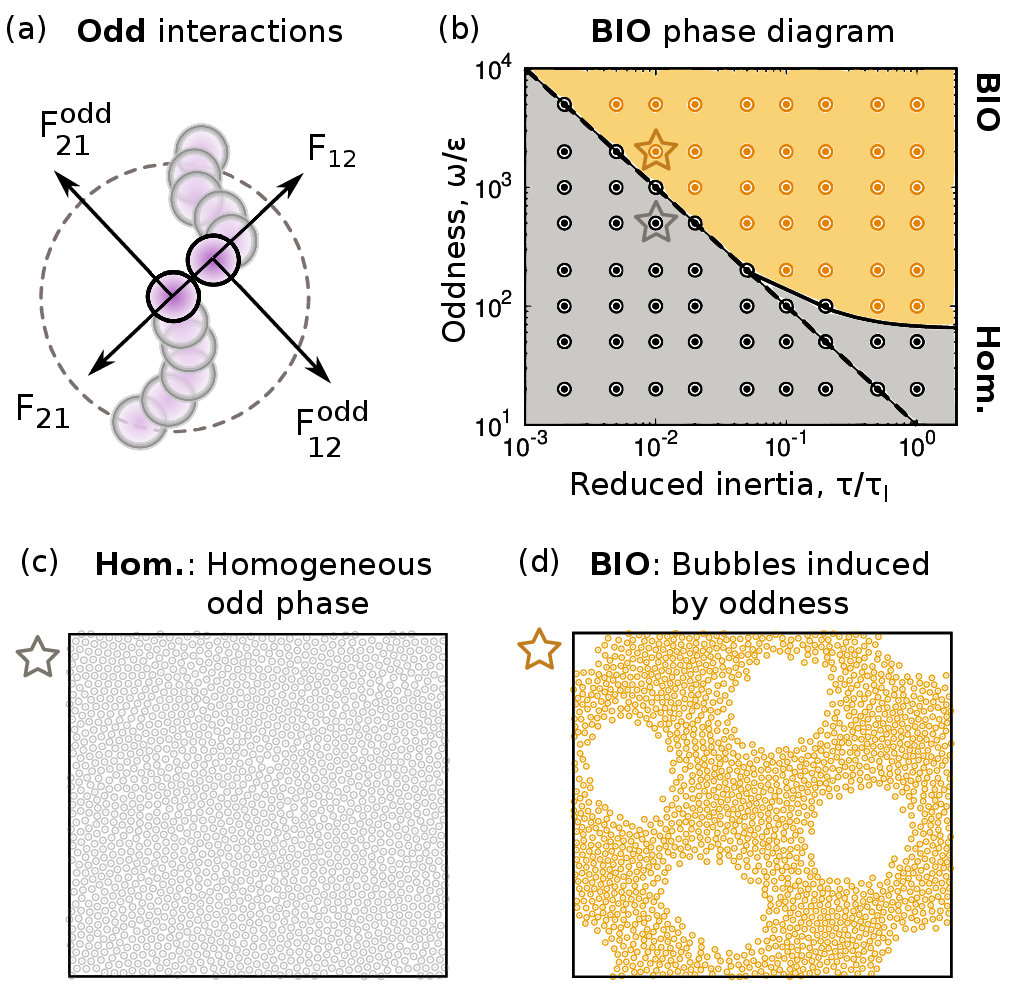}
\caption{
{\textbf{Bubbles induced by odd interactions (BIO).}}
(a): Illustration of a collision between two particles, governed by pure repulsive forces and odd interactions, $F$ and $F^{odd}$.
(b): Phase diagram showing the homogeneous phase (grey) and the BIO phase (yellow).
The dashed line is obtained by plotting the scaling relation between the critical odd parameter and the critical inertial time $\omega^c \sim 1/\tau^c_I$.
(c)-(d): Typical snapshots showing the homogeneous (c) and BIO (d) phases, corresponding to the grey and yellow stars in (a).
The remaining parameters of the simulations are: $T_r=0.2$ and $\Phi=1$.
}\label{fig:Fig1_phasediagram}
\end{figure}

Previous studies have primarily focused on coarse-grained theories where inertia is negligible. In such cases, exceptional points appear in response to mechanical perturbations~\cite{fruchart2023odd}. However, a recent particle-based model for overdamped odd liquids has been proposed~\cite{caporusso2024phase}. It shows that the interplay between attractions and oddness induces different phases, ranging from clustering to chiral liquids and solids, depending on the packing fraction. By contrast, odd systems with purely repulsive interactions, such as spinners, have been poorly explored. Since odd interactions do not conserve energy and generate angular momentum, inertia may not serve as an efficient dissipation mechanism, unlike in equilibrium systems, thus disrupting the stability of the homogeneous phase.
\begin{figure}[!t]
\centering
\includegraphics[width=1\linewidth,keepaspectratio]{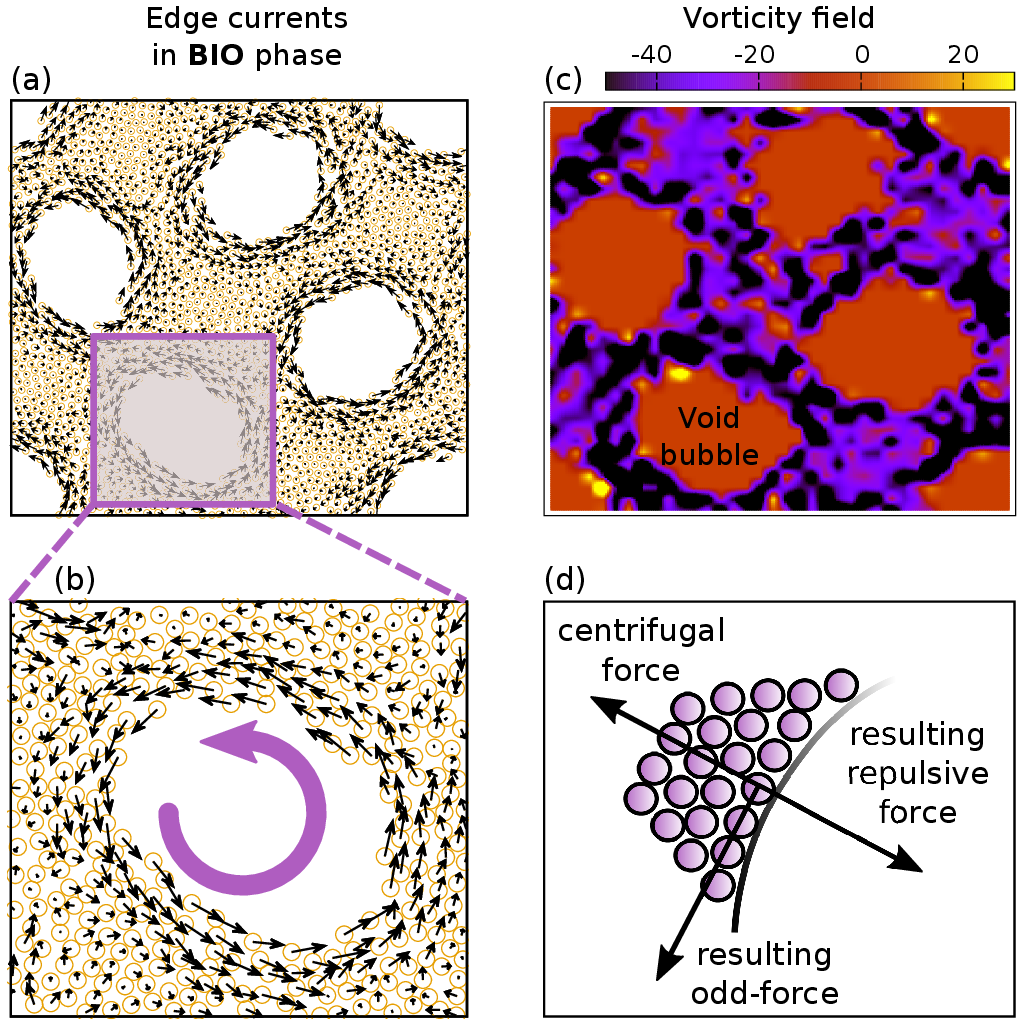}
\caption{
{\textbf{Dynamics of the BIO phase}}.
(a): Zoom for a snapshot in the BIO phase with $\omega/\epsilon=2\times10^{3}$ and $\tau_I/\tau=10^{-2}$, where black arrows represent the particle velocities.
(b): Additional zoom of (a) on a typical bubble with a violet arrow outlining the edge current.
(c): Snapshot for the BIO configuration of (a), reporting the vorticity field as a color gradient. Void regions are characterized by vanishing vorticity (red colors).
(d): Illustration of the mechanism responsible for the bubble stability showing the competition between the standard repulsive force and the centrifugal force due to odd interactions. 
The remaining parameters of the simulations are: $T_r=0.2$ and $\Phi=1$.
}\label{fig:edge_currents}
\end{figure}

Here, we discover that the interplay between odd forces and inertia generates a non-equilibrium phase transition (Fig.~\ref{fig:Fig1_phasediagram}), from a homogeneous to a non-homogeneous phase, characterized by bubbles induced by odd-interactions (BIO).
This phenomenon occurs spontaneously and without the need for attraction.
Furthermore, it cannot be observed in overdamped odd systems, where the homogeneous phase remains stable due to significant dissipation by friction.
In the BIO phase, bubbles are spatially ordered and separated by vortex structures, with edge currents emerging at the bubble surface. 
As the system approaches the transition line from the homogeneous phase to the BIO phase, spatial velocity correlations with analytically predicted oscillations appear, suggesting the emergence of vortex structures in the homogeneous phase as precursors to the phase transition.

We employ a microscopic approach by evolving the particle dynamics with purely repulsive and odd forces.
The dynamics of a particle with mass $m$ are described by an underdamped equation of motion for the particle velocity $\textbf{\textit{v}}_i=\dot{\mathbf{x}}_i$
\begin{equation}
m \dot {\textbf{\textit{v}}}_i=- \gamma \textbf{\textit{v}}_i
+\mathbf{F}_i+\mathbf{F}^{odd}_i + \sqrt{2\gamma T} \,\boldsymbol{\eta}_i \,,
\label{eq:dynamics}
\end{equation}
where $\boldsymbol{\eta}_i$ is a white noise with zero average and unit variance. The constants $\gamma$ and $T$ represent the friction coefficient and the thermal temperature, respectively.
Here, ${\bf F}_i=-\nabla_i U^{tot}$ is a force due to a total potential which can be expressed as $U^{tot}=\frac{1}{2}\sum_j  U(|\mathbf{x}_i - \mathbf{x}_j)$. In the numerical study, we fix $U(r)$ as a Week-Chandler-Andersen potential $U(r)=4\epsilon((\sigma/r)^{12} - (\sigma/r)^{6})+\epsilon$ for $r<2^{1/6}$ and zero otherwise~\cite{andersen1971relationship}. The constant $\sigma$ represents the nominal particle diameter while $\epsilon$ determines the typical energy scale of the interactions.
Furthermore, the dynamics is governed by additional forces, $\mathbf{F}^{odd}_i$, where the superscript $odd$ indicates that this is an odd force given by:
\begin{equation}
\mathbf{F}^{odd}_i = -  \sum_j \nabla_i U^{odd}(|\mathbf{x}_i-\mathbf{x}_j|)\times \mathbf{z}\,.
\end{equation}
Here, $\mathbf{z}$ is the unit vector, normal to the plane of motion.
The pairwise odd force, generated by the function $U^{odd}(r)=\omega (\sigma/r)$, is cut and shifted so that it vanishes for $r>5\sigma$. This shape is inspired by granular experiments involving spinners where odd interactions have been observed~\cite{massana2021arrested}. 
 Importantly, the term $\omega$ sets the strength of odd interactions and represents our key parameter.
 As a consequence, $\mathbf{F}^{odd}_i$  transfers energy orthogonally, i.e. the $x$ component of the gradient of $U^{odd}(r)$  affects the $y$-dynamics (and vice-versa with the opposite sign). The effect of $\mathbf{F}^{odd}_i$ during a collision between two particles is illustrated in Fig.~\ref{fig:Fig1_phasediagram}~(a).
This microscopic, transversal force, previously employed in Ref.~\cite{caporusso2024phase} in combination with the dynamics~\eqref{eq:dynamics}, can be viewed as a microscopic model for the general hydrodynamics picture developed for odd materials~\cite{fruchart2023odd}.

We evolve the dynamics~\eqref{eq:dynamics} in two dimensions for $N$ particles in a box of size $L$ with periodic boundary conditions (See Supplemental Material~\cite{SM} for further details). The packing fraction is $\Phi= N/L^2 \sigma^2 \pi/4 = 1$, so that the system shows a homogeneous dense configuration in the absence of odd interactions.
By using the particle diameter $\sigma$ as a unit of length and $\tau=\sigma \sqrt{m/\epsilon}$ as a unit of time, the system is controlled by three dimensionless parameters: The reduced inertial time, $\tau_I/\tau$ where  $\tau_I=m/\gamma$ is the ratio of the particle mass and friction; The oddness $\omega\tau^2/(\sigma^2 m)=\omega/\epsilon$ which is the ratio between the odd and even energy scales; and finally the reduced temperature $T_{r}= T\gamma \tau^{3}/(\sigma^2 m^2) $, which determines the noise strength compared to the potential energy scale.
The last parameter is kept constant because does not change the phenomena depicted here.

The effects of the oddness $\omega/\epsilon$ and the reduced inertial time $\tau_I/\tau$ are systematically analyzed in a phase diagram at a fixed packing fraction (Fig. \ref{fig:Fig1_phasediagram}~(b)). For small values of $\omega/\epsilon$ and $\tau_I/ \tau$, the system shows a homogeneous phase (Fig.~\ref{fig:Fig1_phasediagram}~(c)) as in a pure repulsive thermal system.
For large values of $\omega/\epsilon$ and $\tau_I/ \tau$, the homogeneous phase becomes unstable, leading to the formation of circular bubbles - regions devoid of particles - that are periodically arranged in space (Fig. 1(d)). This stable density inhomogeneity occurs despite the absence of attractive forces and is a non-equilibrium phenomenon induced purely by odd interactions.

\begin{figure}[!t]
\centering
\includegraphics[width=1\linewidth,keepaspectratio]{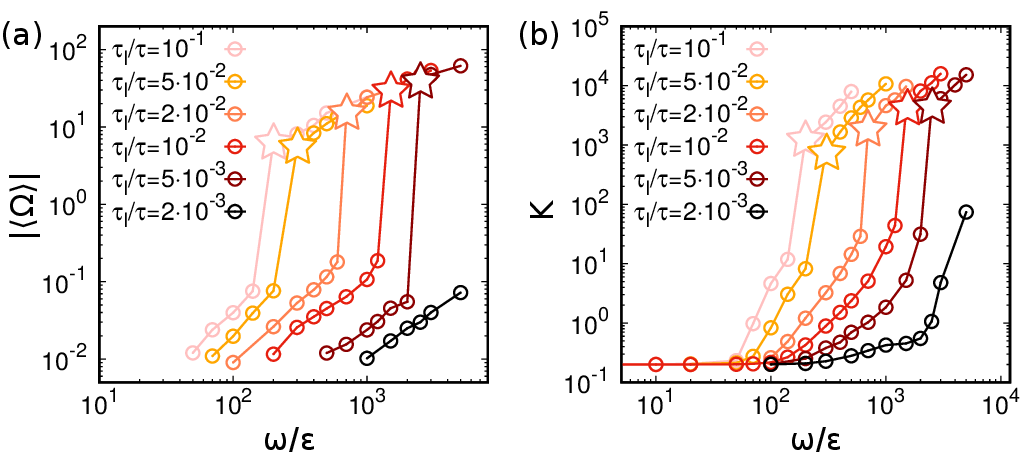}
\caption{
{\textbf{Order parameters.}}
(a): Modulus of the total average vorticity $|\langle \Omega\rangle|$ as a function of the odness $\omega/\epsilon$.
(b): Kinetic energy $m\langle \mathbf{v}^2\rangle/2$ as a function of the oddness $\omega/\epsilon$.
Both in (a) and (b), the analysis is performed for different values of the reduced inertia $\tau_I/\tau$ while stars denote the first $\omega/\epsilon$ value where the BIO transition takes place at a given $\tau_I/\tau$.
The remaining parameters of the simulations are: $T_r=0.2$ and $\Phi=1$.
}\label{fig:Fig1_kinetic}
\end{figure}

The particles placed at the bubble surface collectively rotate creating edge currents (Fig.~\ref{fig:edge_currents}~(a)-(b)).
This effect has been previously observed in experiments~\cite{soni2019odd, massana2021arrested} and numerical studies~\cite{caporusso2024phase}.
Furthermore, here, bubbles are separated by giant vortices, as emerged by plotting the local vorticity field (Fig.~\ref{fig:edge_currents}~(c)), $\Omega(\mathbf{x}) =  \left( \partial_x v_y(\mathbf{x}) - \partial_y v_x(\mathbf{x}) \right)$, where $v_x$ and $v_y$ are the two Cartesian components of the velocity field.
We expect a large negative global vorticity in BIO configurations and a very small one in the homogeneous phase.
Thus, $|\langle \Omega\rangle|$ serves as the order parameter for the BIO phase, as shown in Fig.~\ref{fig:Fig1_kinetic}~(a) as a function of the oddness $\omega/\epsilon$ for different values of the reduced inertia $\tau_I/\tau$.
When $\omega/\epsilon$ approaches the transition line, $|\langle\Omega\rangle|$ displays a sharp jump: it takes small values close to zero for homogeneous configurations while it is characterized by a large value $|\langle\Omega\rangle| \gg 1$ in the BIO phase.
This phenomenon occurs because the energy injected by odd forces cannot be efficiently dissipated by friction.
Therefore, the total kinetic energy of the system $K=m \langle \mathbf{v}^2\rangle/2$ displays a giant increase as a function of $\omega/\epsilon$ at the transition (Fig.~\ref{fig:Fig1_kinetic}~(b)) reaching a value larger than the thermal temperature $T$, as checked for different values of the reduced inertia $\tau_I/\tau$.

In the BIO phase, steady-state bubbles with edge currents can be observed even in the absence of attractive interactions.
The bubble stability can be deduced by combining an energetic and a mechanical argument.
In a homogeneous phase, repulsive forces among particles mutually balance as also odd forces, resulting in no net average force at any specific location due to symmetry.
By contrast, in the inhomogeneous phase, a particle on the bubble surface experiences unbalanced forces from the remaining particles. Repulsive forces push the particle towards the bubble's interior, exerting a net inward force normal to the bubble surface. This net force is similar to equilibrium pressure, which tends to suppress bubbles, and scales as $|\mathbf{F}^{rep}| \sim \epsilon/\sigma$.
At the interface, odd interactions due to the remaining particles are also unbalanced, creating an additional net force tangential to the bubble surface, scaling as
$|\mathbf{F}^t| \sim \omega/\sigma$.
This tangential force induces a persistent rotation of the particle on the bubble surface and is responsible for the numerically observed edge currents.
The rotating motion of the particle due to odd forces generates a centrifugal force pointing outward from the bubble surface's normal direction. Its modulus is given by $|\mathbf{F}^{cen}|=m v^2_t/ R$, where $v_t$ is the tangential velocity induced by $\mathbf{F}^t$ and $R$ is the bubble radius which is $\approx \sigma$ at the BIO transition point.
The BIO transition occurs when mechanical and dissipative force balance, i.e. when the tangential force due to odd interactions is equal to the viscous force, $|\mathbf{F}^t|\approx\gamma |\mathbf{v}_t|$, and when the odd-induced centrifugal force is balanced by the repulsive force due to the remaining particles, $m |\mathbf{F}^t|/\gamma^2 \approx |\mathbf{F}^{rep}|$.
Using the expressions for $|\mathbf{F}^{rep}|$ and $|\mathbf{F}^{t}|$, we can predict the scaling for oddness and inertia at the BIO transition line:
\begin{equation}
\label{eq:scaling}
\omega_c \sim \frac{\sqrt{m}}{\tau^c_I} \,.
\end{equation}
The scaling~\eqref{eq:scaling} is numerically confirmed in the phase diagram (Fig.~\ref{fig:Fig1_phasediagram}~(b))
and is violated only for large inertia values $\tau_I/ \tau$ where the homogeneous phase becomes more stable.

\begin{figure}[!t]
\centering
\includegraphics[width=1\linewidth,keepaspectratio]{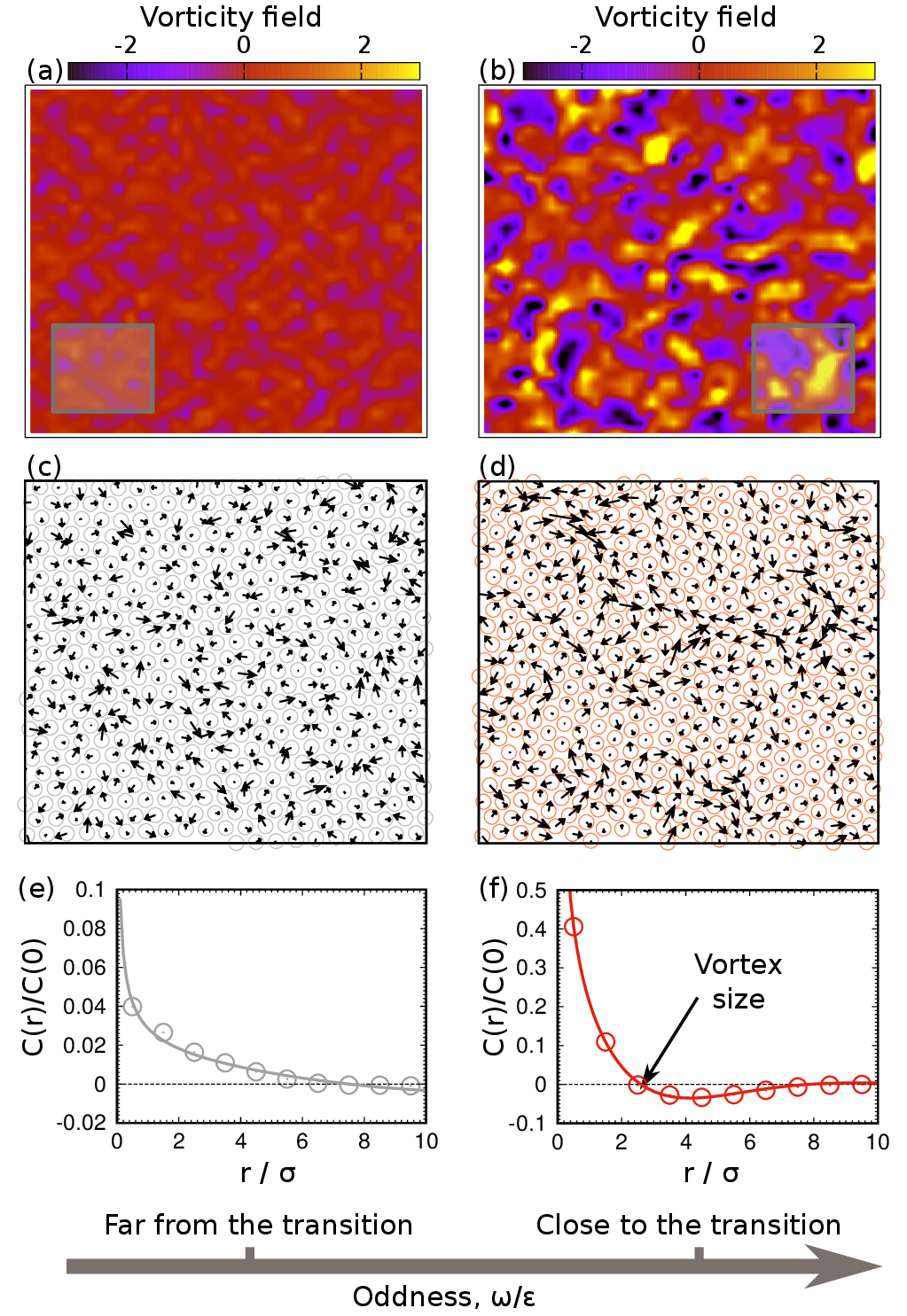}
\caption{
{\textbf{Dynamical properties below the BIO transition}}.
(a)-(b): Snapshots of the homogeneous odd phase showing the vorticity field as a color gradient, for $\omega/\epsilon=10^2$ (far from the transition line) and $\omega/\epsilon=5\times10^2$ (close to the transition line).
(c)-(d): Zooms of panels (a) and (b), respectively, corresponding to the square regions in (a) and (b). In these panels, black vectors represent the particle velocities.
(e)-(f): Spatial velocity correlations for $\omega/\epsilon=10^2$ and $\omega/\epsilon=5\times10^2$, respectively. Here, dashed black lines are eye-guides to mark zero, while solid colored lines are obtained by fitting the function $a e^{-r/b} cos(r c)/r^{1/2}$, i.e. the inverse Fourier Transform of Eq.~\eqref{eq:dyn_corr}, with $a$, $b$ and $c$ as fitting parameters.
The remaining parameters of the simulations are: $T_r=0.2$ and $\Phi=1$.
}\label{fig:3}
\end{figure}

In the homogeneous phase, the system exhibits non-equilibrium dynamic structures that vanish when the system is sufficiently far from the transition line to BIO. These structures can be visualized by calculating the local vorticity field, $\Omega(\mathbf{x})$. When the oddness is much smaller than the critical value, $\omega \ll \omega_c$, $\Omega(\mathbf{x})$ is nearly zero everywhere (Fig.\ref{fig:3}~(a)), and the velocities point in random directions, as is typical in equilibrium thermal systems (Fig.\ref{fig:3}~(c)). However, when the oddness approaches the critical value from below, $\omega \approx \omega_c$, the vorticity field displays spatial patterns with alternating regions of positive and negative vorticity, with vortices and antivortices (Fig.\ref{fig:3}~(b)). Consequently, the particle velocities also form spatial patterns (Fig.\ref{fig:3}~(d)), showing a degree of alignment and vortex-like structures even in the absence of alignment interactions.

This phenomenon is a clear signature of the transition, analogous to equilibrium Ising models where spatial correlations are precursors of  critical phenomena. To quantify this effect, we have measured the spatial velocity correlations, $\langle \mathbf{v}(r) \cdot\mathbf{v}(0)\rangle$, which also emerge without alignment interactions. Specifically, $\langle \mathbf{v}(r) \cdot\mathbf{v}(0)\rangle$ is negligible far from the transition line, $\omega \ll \omega_c$ (Fig.\ref{fig:3}(e)). In contrast, near the BIO-transition line at $\omega \approx \omega_c$, velocity correlations exhibit a spatial profile that crosses zero and oscillates (Fig.\ref{fig:3}(f)). This behavior indicates the presence of vortices, entirely induced by odd interactions.


The oscillating profile for the spatial velocity correlations is analytically predicted for a dense system governed by odd interactions (See SM~\cite{SM} for the calculation details).
The profile in Fourier space reads:
\begin{equation}
\label{eq:dyn_corr}
\langle \hat{\mathbf{v}}(\mathbf{q})\cdot\hat{\mathbf{v}}(-\mathbf{q}) \rangle=2\frac{T}{m}\Bigl( 1+B(\bq)\Bigr) \,,
\end{equation}
where $\mathbf{q}$ is the wave vector and $\hat{\mathbf{v}}(\mathbf{q})$ the Fourier transform of the velocity field.
These correlations consist of two contributions: (i) a constant term, which does not give rise to any spatial structure  as in equilibrium; (ii) a non-equilibrium term,
$B(\mathbf{q})$, which depends on $\mathbf{q}$. This term introduces a spatial structure in the velocity field that disappears when odd interactions vanish
($\omega=0$).
We present  the expression for $B(\mathbf{q})$ close to the transition line (see~\cite{SM} for the full prediction):
\begin{equation}
B(\mathbf{q})\propto\frac{1}{a+ b(\bq-{\bf q}^*)^2}\,,
\label{lorenzianabrillouin}
\end{equation}
where $\mathbf{q}^*$ is a non-vanishing wave vector while $a$ and $b$ are two positive constants (see~\cite{SM} for their expression).
Equation~\eqref{lorenzianabrillouin} defines the effective correlation length, $\xi$, of the spatial velocity correlations
as $\xi=\sqrt{b/a}$, scaling as
\begin{equation}
\label{eq:xi}
\xi^2\sim\frac{\beta_2 \tau_I^2\frac{C_1^2 }{ m C_0 }}{1-\beta_1 \tau_I^2 C_1^2/m C_0} \,.
\end{equation}
In this expression, $C_0$ denotes the standard elastic constant induced by repulsive interactions and
and $C_1$  the odd elastic constant, i.e. the non-diagonal component of the elastic tensor.  Since the odd constant scales as $C_1 \sim \omega$, the correlation length $\xi$ is an increasing function of the oddness $\omega$, as shown by Eq.~\eqref{eq:xi}. Its divergence implies the existence of a dynamical instability.
This allows us to identify critical values of oddness and inertia that are consistent with the scaling $\omega_c \sim \sqrt{m}/\tau^c_I$ numerically observed for the BIO transition.

Finally, we note that the expression~\eqref{lorenzianabrillouin} does not exhibit the standard Ornstein-Zernike profile typical of phase transitions in equilibrium~\cite{ma2018modern} or the spatial patterns observed in dense self-propelled particles~\cite{caprini2020spontaneous, caprini2021spatial, szamel2021long, marconi2021hydrodynamics, keta2022disordered}. Indeed, in our system, the transition occurs for a non-vanishing wavelength vector, $\mathbf{q}^*$ (see SM~\cite{SM} for its definition).
By applying the inverse Fourier transform to Eq.~\eqref{lorenzianabrillouin} (see SM~\cite{SM} for details), we observe that spatial velocity correlations are not only characterized by an exponential decrease, determined by the typical length $\xi$, but also by a sinusoidal modulation in space. This is the profile fitted in Fig.~\ref{fig:3}~(f) which agrees with our numerical findings.
The oscillations in spatial velocity correlations also explain the vortex structures observed numerically near the transition point: the first zero crossing of $\langle \mathbf{v}(r) \cdot \mathbf{v}(0) \rangle$ determines the typical vortex size.

Our findings reveal a phase transition spontaneously occurring far from equilibrium, which can be experimentally investigated in dense systems of granular spinners. The discovered pattern formation could have applications in metamaterials~\cite{bertoldi2017flexible, veenstra2024non} and odd, chiral liquids. In these cases, odd forces generate an effective surface tension that counteracts standard pressure and creates stable bubbles even in the absence of attractions.
Deriving a non-equilibrium scalar field theory to reproduce these effects may represent a challenging future research step.




\bibliographystyle{apsrev4-1}

\bibliography{odd.bib}

\end{document}